# 2DEGs at Perovskite Interfaces between $KTaO_3$ or $KNbO_3$ and Stannates

Xiaofeng Fan[1], Weitao Zheng[1], Xin Chen[2], David J. Singh[2]*

1 College of Materials Science and Engineering, Jilin University, Changchun, People's Republic of China, 2 Materials Science and Technology Division, Oak Ridge National Laboratory, Oak Ridge, Tennessee, United States of America

**Abstract**

We report density functional studies of electron rich interfaces between $KTaO_3$ or $KNbO_3$ and $CaSnO_3$ or $ZnSnO_3$ and in particular the nature of the interfacial electron gasses that can be formed. We find that depending on the details these may occur on either the transition metal or stannate sides of the interface and in the later case can be shifted away from the interface by ferroelectricity. We also present calculations for bulk $KNbO_3$, $KTaO_3$, $CaSnO_3$, $BaSnO_3$ and $ZnSnO_3$, showing the different transport and optical properties that may be expected on the two sides of such interfaces. The results suggest that these interfaces may display a wide range of behaviors depending on conditions, and in particular the interplay with ferroelectricity suggests that electrical control of these properties may be possible.

**Funding:** Work at ORNL was supported by the Department of Energy, Basic Energy Sciences, Materials Sciences and Engineering Division. The funders had no role in study design, data collection and analysis, decision to publish, or preparation of the manuscript.

**Competing Interests:** The authors have declared that no competing interests exist.

* E-mail: singhdj@ornl.gov

## Introduction

Oxide electronics is an exciting area that is also of great potential technological importance. This arises from the wide range of properties available in oxides, even staying within a given structural family, such as perovskite. These include high conductivity metallic behavior as well as ferroelectricity and ferromagnetism. These functionalities may lead to new devices that are difficult to construct using conventional semiconductors. Key parameters for the conducting channels of electronic devices are the conductivity, the carrier velocity and the mobility. Furthermore, discoveries of new emergent phenomena, particular metallic 2 dimensional electron gasses (2DEGs) at certain oxide interfaces have led to renewed interest. [1] While initially it was thought that the 2DEGs represented a particular manifestation of physics of strongly correlated electrons at the interface between Mott and band-insulators, it is now recognized that it is a much more general phenomena related to carriers being forced into electronic states at the interfaces by electric fields associated with charge imbalance (shown e.g. by experiments on interfaces between $LaAlO_3$ and $SrTiO_3$) [2].

The main challenge at present is to find materials systems and methods for producing controlled, usually high carrier density 2DEGs, with high mobility and a variety of functional behaviors. To date almost all work has focused on mixed valent transition element based perovskites, mainly titanates. One explanation for the success with titanates is that Ti readily takes valences between $Ti^{+4}$ and $Ti^{+3}$ in bulk oxides and that such systems are often metallic as in doped $SrTiO_3$, which is a superconductor. [3] The implication is that chemically stable, metallic electron doped titanates can exist. This in contrast to oxides based on metals that do not have mixed valence states as in those cases compensating defects may be expected instead of metallic conductivity. In the perovskite family, besides titanates, both niobates and tantalates have been investigated theoretically, with promising results. [4] Furthermore, electrostatically doped $KTaO_3$ at the interface with an ionic liquid has been shown to be superconducting, [5] suggesting that novel physics may also be found at other charged $KTaO_3$ interfaces. However, there has been very little work with non-transition metal oxides.

Here we report studies of n-type interfaces based on combinations of $K(Ta,Nb)O_3$ and $(Ca,Zn)SnO_3$. The motivation is as follows: (1) The oxides on both sides of the interface have mixed valences and can become highly conducting with electron doping and are naturally n-type materials again suggesting the possibility of maintaining electron carriers (i.e. against compensating defects); (2) $KNbO_3$ is a well known ferroelectric, where the ferroelectric transition can be tuned down to 0 K by alloying with Ta and similarly $ZnSnO_3$ is a ferroelectric [6–10] - this provides the possibility of making 2DEGs in proximity to ferroelectric materials and also provides an opportunity for tuning the dielectric constants over a wide range. Note that recent work connects the presence of a high dielectric constant, as is the case in $SrTiO_3$, with the occurrence of high mobility due to screening of scattering centers); [11,12] (3) since both sides of the interface are potentially conducting, it may be that one can arrange the compositions so that one can switch the 2DEG from the Ta/Nb side to the Sn side with e.g. electric field thereby perhaps drastically changing functional properties; and finally (4) it is of interest to explore what properties may be expected for a 2DEG based on a $s$ electron system such as $ZnSnO_3$ as opposed to the $d$-electron systems studied so far. In particular, $s$ electron materials tend to have much wider bands with potential implications both for the mobility and achievable conductivities.

We note that there are a number of highly conducting but transparent oxide materials, such as Sn doped $In_2O_3$ (ITO) and the recently discovered doped $BaSnO_3$. [13–18] The high conductivity of doped n-type $BaSnO_3$ is indicative of a potential for highly metallic 2DEGs at interfaces of $Sn^{4+}$ perovskites. We note that n-type $BaSnO_3$ was prepared early on with doping by Sb and that its band structure was studied emphasizing the s-electron character of the conduction band [19,20].

The lattice parameter of cubic $BaSnO_3$, $a = 4.116$ Å, [21–23] is larger than most available transition metal perovskites that could be used as substrates for growth. The ~3% compressive strain that would be imposed in growth of $BaSnO_3$ on e.g. $KTaO_3$ substrates is within the range that is possible for perovskite growth; in fact epitaxial growth on $SrTiO_3$ has been reported. [16,24] This includes very high mobility (up to 70 $cm^2V^{-1}s^{-1}$) material. [16] Here we focus on $CaSnO_3$ and $ZnSnO_3$, which have lattice parameters that are more compatible with perovskite $KNbO_3$ and $KTaO_3$.

$CaSnO_3$ occurs in an orthorhombic $GdFeO_3$ perovskite type structure characteristic of low tolerance factor materials, [21] while, as mentioned, the related compound $ZnSnO_3$ is a $LiNbO_3$ type ferroelectric. The $LiNbO_3$ structure can be viewed as a highly distorted perovskite structure where the octahedra tilt until blocked by ionic repulsions among the B-site (Nb) and O ions. This still leaves the A-site ions in sites that are too large based on their ionic radii after the tilt. These A-site ions then off-center to obtain a more suitable coordination.

While initially found as a phase formed by ion exchange [6] or high pressure, [7] $ZnSnO_3$ can be readily grown by hydrothermal and carbon evaporation methods, [25–29] and importantly as heteroepitaxial films on perovskite substrates. [9] $KTaO_3$ is a cubic perovskite that is close to ferroelectricity but is not ferroelectric. This absence of a ferroelectric state down to the lowest temperatures is the case in experiment and also in (zero temperature) density functional calculations. [30–32] The lattice parameter of $KTaO_3$ is 3.99 Å. This is a reasonable match for $CaSnO_3$, which has orthorhombic lattice parameters of 5.681 Å, 7.906 Å and 5.532 Å (i.e. $\sqrt{2} \times 4.017$ Å, $2 \times 3.953$ Å and $\sqrt{2} \times 3.912$ Å). $KTaO_3$ is an indirect band gap material with an experimental gap of ~3.6 eV [30,33].

$KNbO_3$ has a smaller gap of ~3 eV. [34] This difference implies a lower position for the transition metal d bands, and greater covalency, which has been related to the different ferroelectric properties of these two materials. [32] $BaSnO_3$, which is reported to have a cubic perovskite structure at ambient temperature with lattice parameter $a = 4.116$ Å and as mentioned is be an excellent n-type transparent conductor with doping.

## Results

### Bulk Compounds

We start with the electronic structure of the bulk compounds. These were based on the experimental cubic structures for $BaSnO_3$ and $KTaO_3$, 4.116 Å and 3.99 Å, respectively. For $CaSnO_3$ we used the experimental orthorhombic lattice parameters (spacegroup 62, Pnma, $a = 5.681$ Å, $b = 7.906$ Å, $c = 5.532$ Å), and relaxed the internal atomic positions using the PBE GGA. Similarly, we used the experimental lattice parameters and spacegroup for $KNbO_3$ (R3m, pseudocubic $a = 4.016$ Å) and $ZnSnO_3$ (R3c, hexagonal setting, $a = 5.2622$ Å, $c = 14.0026$ Å), and relaxed the internal atomic positions consistent with symmetry with the PBE GGA and then calculated electronic structure with the TB-mBJ potential functional.

The band structure and calculated optical absorption of cubic $BaSnO_3$ are shown in Figs. 1 and 2, respectively. The TB-mBJ band structure of $BaSnO_3$ is in close agreement with the recent hybrid functional calculations of Liu and co-workers, [35] including the indirect band gap, its value and the structure of the conduction bands. Kim and co-workers observed that while the TB-mBJ functional gives band gaps of III-V semiconductors in good accord with experiment it tends to underestimate band widths relative to hybrid functional and GW calculations. In $BaSnO_3$ we obtain a dispersion of the lowest conduction band from $\Gamma$-X of 3.7 eV as compared to ~4.3 eV in the hybrid functional calculations of Liu and co-workers, [35] which amounts to a difference of ~15%. It will be of interest to compare the structure of measured optical spectra with the present calculations (Fig. 2) to determine the band widths. Critical point analysis of ellipsometry data may also be helpful for this.

As expected, the valence bands have O 2p character while the conduction band minimum (CBM) has Sn s character. The calculated refractive index of $BaSnO_3$ is 1.9 at 2 eV. This is approximately 10% smaller than the experimental single crystal value reported by Stanislavchuk and co-workers. [36] The band structure of $KTaO_3$ is shown in Fig. 3 The structure of the CBM in the different compounds reflects the different orbital composition. In $KTaO_3$ and $KNbO_3$ the CBM has transition metal $t_{2g}$ d character, i.e. in the absence of spin orbit splitting it is three-fold degenerate at $\Gamma$ for the cubic compound $KTaO_3$ and derives from the $d_{xz}$, $d_{yz}$ and $d_{xy}$ orbitals. These lead to nearly two dimensional bands. This is a consequence of the fact that with the perovskite bonding topology the $d_{xy}$ band has much reduced dispersion along $k_z$ since this dispersion would arise from hopping through the O atom along the z direction, which is not allowed by symmetry, and similarly for the other two $t_{2g}$ orbitals. This structure is reflected in the band structures of interfacial systems, such as those based on $SrTiO_3$ [37].

In the case of $KTaO_3$, the spin-orbit interaction splits these into a lower lying two fold degenerate band that retains this structure and a higher lying single degenerate band that is approximately isotropic. Conventionally, these are described as $j = 3/2$ and $j = 1/2$ bands, respectively, based on the analogy of the $t_{2g}$ manifold with an effective p manifold. Fig. 4 illustrates this through an isoenergy surface 0.1 eV above the CBM. In contrast the CBM of the Sn compounds comes from the Sn $s$ orbital leading to a strongly dispersive isotropic band.

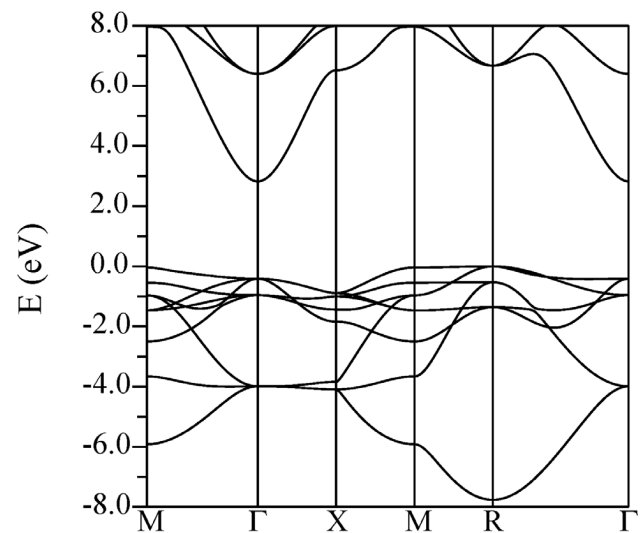

**Figure 1. Calculated band structure of cubic $BaSnO_3$ using the TB-mBJ potential.**
doi:10.1371/journal.pone.0091423.g001

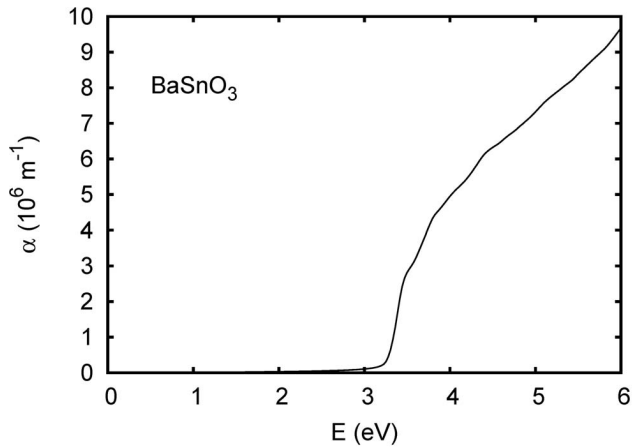

**Figure 2. Calculated absorption spectrum of BaSnO$_3$ using the TB-mBJ potential.** A Lorentzian broadening of 0.025 eV was applied.
doi:10.1371/journal.pone.0091423.g002

This has consequences for transport. Within Boltzmann transport theory, the conductivity, $\sigma = ne\mu$ in the degenerate regime at low temperatures is proportional to $N(E_F) <v_F^2> \tau$, where $N(E_F)$ is the density of states at the Fermi energy, $E_F$, $<v_F^2>$ is the average square Fermi velocity in the direction of transport and $\tau$ is the effective inverse scattering rate (at finite temperature the formulas are similar, but involve integration with Fermi functions). For an isotropic parabolic band this can be expressed in terms of the effective mass, $m^*$, $\sigma \propto (n/m)\tau$. The low temperature (Mott formula) Seebeck coefficient, $S(T) = \pi^2 k_B^2 T(d\sigma/d)/(3e\sigma)$ is proportional to $-Tm/n^{2/3}$ in the single parabolic band case and such expressions involving $n$ and $m$ can be obtained for other transport coefficients as well. In deriving these expressions, the effective mass takes two generic roles: (1) in the band velocity, $v = d_k/dk$ and (2) in the number of carriers, which is given by twice (due to spin) the volume of the Fermi surface, $8\pi k_F^3/3$, $k_F = (2mE_F)^{1/2}$, where $E_F$ is the Fermi energy. At low $T$, $S(T) \propto T/E_F$. For the anisotropic case (see Refs. [38] and [39] for a discussion of the maximally anisotropic case), where the inverse effective mass is a rank two tensor, $\sigma$ in the light mass direction at a given $E_F$ enhanced because the carrier concentration $n$ will be higher than for an isotropic system with light mass in all directions, and accordingly the conductivity will be higher if the scattering rate is not proportionately enhanced. One the other hand $S(T)$ will be the same as in the isotropic light band case, with the same Fermi energy. Conversely, for the same carrier density, $n$, the isotropic light band system will have a higher $E_F$ than the anisotropic case and therefore a reduced magnitude of $S(T)$. This effect will be particularly pronounced at high carrier densities.

This was discussed previously in the context of the thermoelectric properties of SrTiO$_3$/LaAlO$_3$ interfaces by Filippetti and co-workers, [40] who pointed out that little is gained in the thermopower from making the electronic structure of SrTiO$_3$ two dimensional through the interface, since one already has the effect of a strong anisotropy in the individual bands. It should be noted that the band degeneracy of three also reduces $E_F$ for a given carrier concentration, enhancing $S(T)$ as was noted by Usui and co-workers. [41] We observe that the band structure KTaO$_3$

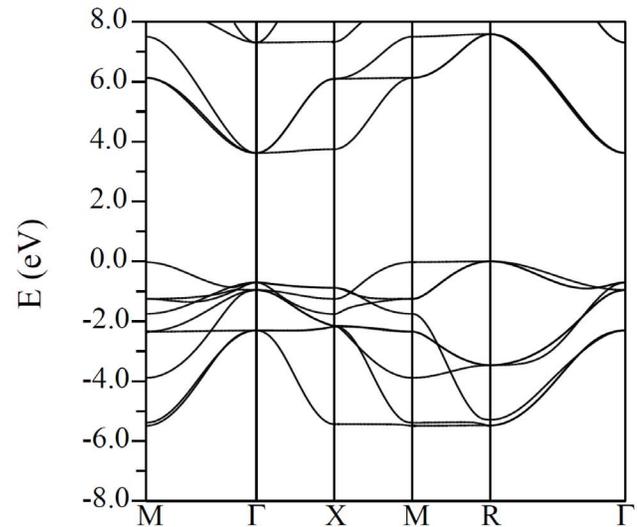

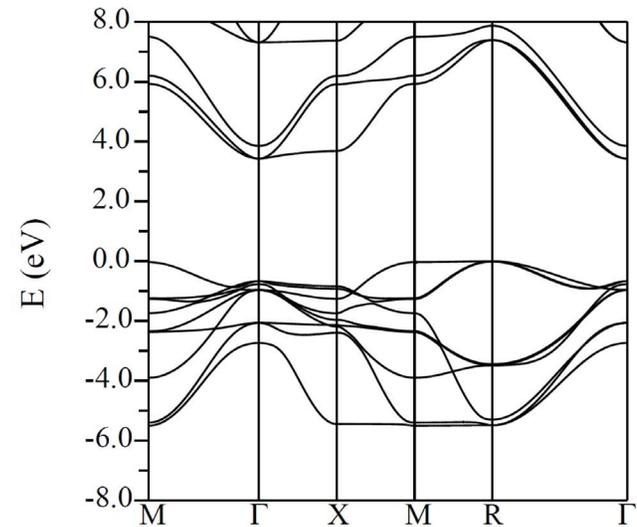

**Figure 3. Calculated band structure of cubic KTaO$_3$ using the TB-mBJ potential, with a scalar relativistic approximation (top) and including spin orbit (bottom).** Note the spin orbit induced splitting of the $t_{2g}$ derived conduction band minimum into a two fold degenerate lower lying effective $j = 3/2$ band and a higher lying single degenerate effective $j = 1/2$ band (considering the $t_{2g}$ manifold as an effective $p$ manifold).
doi:10.1371/journal.pone.0091423.g003

is mostly favorable from a thermoelectric point of view because of the anisotropy (Fig. 4).

BaSnO$_3$ is in the opposite limit, with a single Sn $s$ derived conduction band. In this respect it is like ZnO from a transport point of view, and may be expected to show transport behavior consistent with single isotropic parabolic band formulas. Thus, although all materials considered here are perovskite oxides, one can expect very different transport behavior depending on which side of the interface the electron gas resides. On the KNbO$_3$/KTaO$_3$ side, one expects a lower $E_F$ electron gas with degenerate anisotropic bands, while on the ZnSnO$_3$/CaSnO$_3$ side on expects a high mobility isotropic electron gas with higher $E_F$ for a given carrier density. To better illustrate the differences in relation to thermoelectricity, we show in Fig. 5 the calculated 300 K thermopower as a function of carrier concentration for BaSnO$_3$ in comparison with that of KTaO$_3$. This was obtained within Boltzmann theory using the constant scattering time approximation, [42] via the BoltzTraP code, [43] as in our previous work on thermoelectric materials. [44,45] As may be seen large values are obtained for KTaO$_3$, consistent with the earlier work of Usui and co-workers [41] and with experiment, [46] but BaSnO$_3$ shows much smaller values at similar carrier concentrations even though the ratio of the effective masses for the conductivity is not so large. As expected, the effect is most prominent at the highest carrier densities.

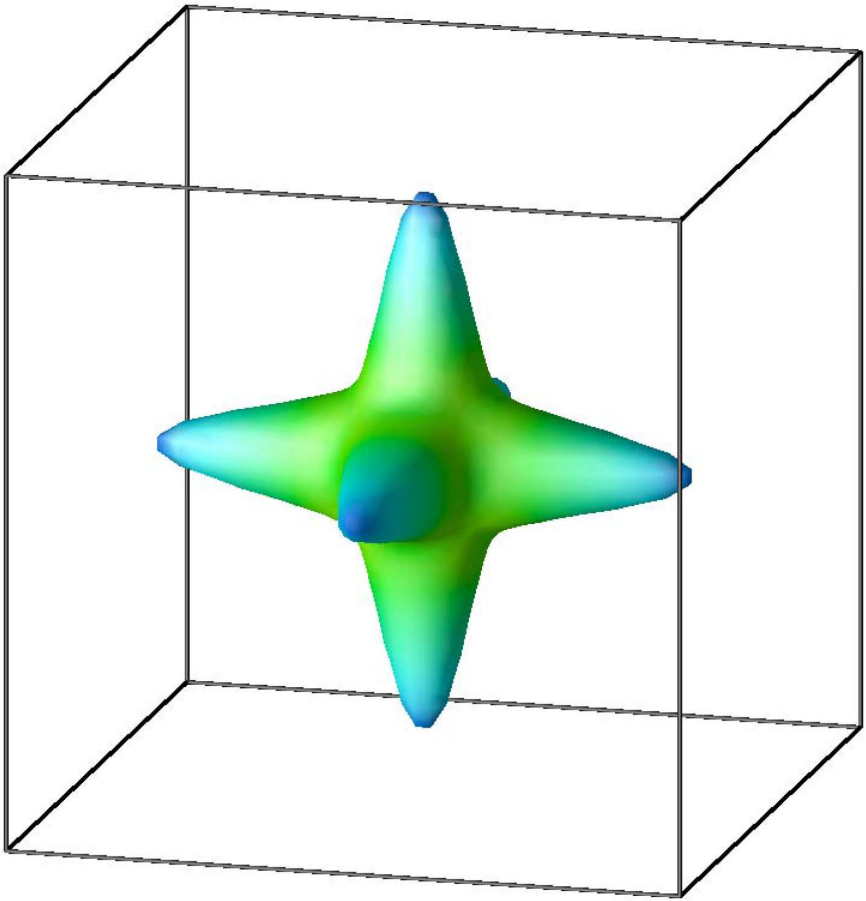

**Figure 4. Outermost isoenergy surface for KTaO$_3$ above the CBM.** There is in addition one inner surfaces (not visible). Note the pronounced non-spherical shape.
doi:10.1371/journal.pone.0091423.g004

Turning to the optical absorption, it is noteworthy that the magnitude of the absorption in BaSnO$_3$ remains rather low at energies above the onset of direct transitions. For example, even at 6 eV (~3 eV above the onset of direct transitions) $\alpha$ is below $10^7$ m$^{-1}$, in contrast to typical transition metal oxides, where there are strong charge transfer excitations and less dispersive conduction bands. Ferroelectric ZnSnO$_3$ (Fig. 6) and orthorhombic CaSnO$_3$ (Fig. 7) show stronger absorption above the onset, although still weaker than typical charge transfer gap transition metal oxides. This is illustrated by the calculated absorption spectrum of KNbO$_3$, shown in Fig. 8. Wiesendanger reported a strong onset of optical absorption at ~4 eV in ferroelectric tetragonal KNbO$_3$, [47] in qualitative accord with the present results.

We obtained a band gap of 3.43 eV for KTaO$_3$, including spin-orbit, in reasonable accord with the experimental value of ~3.6 eV. This calculation included spin orbit. This significantly affects the band structure due to the high atomic number of the 5$d$ element, Ta ($Z=73$), as has recently been emphasized in the context of 2DEGs at KTaO$_3$ interfaces [48].

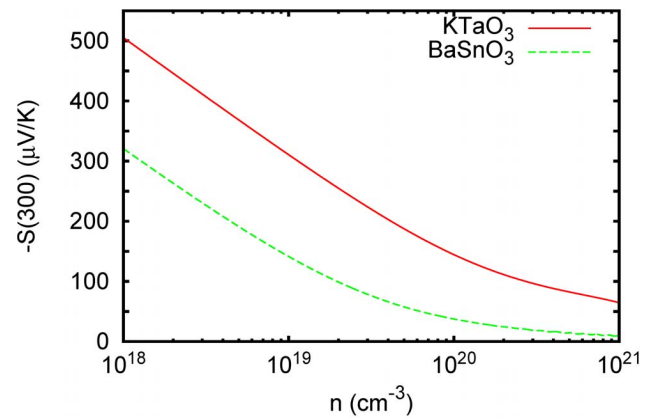

**Figure 5. Calculated Seebeck coefficient of n-type KTaO$_3$ and BaSnO$_3$ as a function of carrier concentration at 300 K as obtained within Boltzmann transport theory with the constant scattering time approximation.** The underlying electronic structures are those obtained using the TB-mBJ potential, and in the case of KTaO$_3$ includes spin orbit.
doi:10.1371/journal.pone.0091423.g005

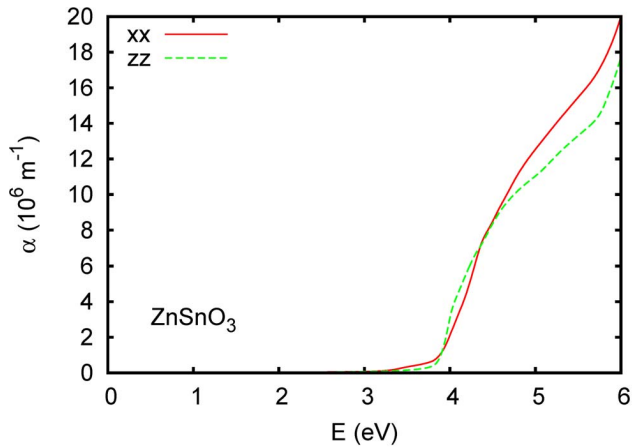

**Figure 6. Calculated absorption spectrum of ZnSnO$_3$ using the TB-mBJ potential.** Here zz is for polarization along the rhombohedral axis and xx is perpendicular. A Lorentzian broadening of 0.025 eV was applied.
doi:10.1371/journal.pone.0091423.g006

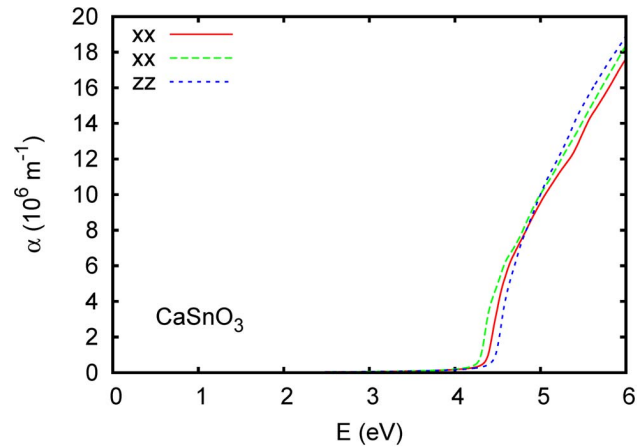

**Figure 7. Calculated absorption spectrum of CaSnO$_3$ using the TB-mBJ potential.** The Cartesian directions are along the crystallographic $a$, $b$ and $c$ orthorhombic lattice parameters. A Lorentzian broadening of 0.025 eV was applied.
doi:10.1371/journal.pone.0091423.g007

Without spin orbit (i.e. scalar relativistic) a gap of 3.61 eV is obtained indicating a non-negligible shift of 0.18 eV when spin orbit is included. We also did a calculation for KTaO$_3$ based on the self-consistent charge density for a scalar relativistic, including spin orbit for the band structure. This non-self-consistent spin orbit calculation yielded a band gap of 3.45 eV, i.e. practically the same as the self consistent calculation. We include spin orbit in the calculations for the interfaces using this approach, i.e. in the final calculation of the electronic structure based on the charge density from self-consistent scalar relativistic calculations.

For BaSnO$_3$ we obtain an indirect gap of 2.82 eV, also in accord with experimental reports. For CaSnO$_3$, we obtain a substantially larger value of 4.17 eV. This reflects distortion of the very dispersive Sn s derived conduction bands. The very strong change in the band gap is remarkable and implies strong deformation potentials. It will be of considerable interest therefore to perform low temperature studies of doped n-type perovskite $A$SnO$_3$, $A$ = Ca,Sr,Ba looking for superconductivity. In order to better show the origin of the band gap change with structure we did calculations for BaSnO$_3$ both as a function of octahedral rotation and as a function of lattice parameter. For the rotation we considered a frozen in $R$-point phonon consisting of alternating (G-type) rotations about a [001] direction. With the experimental lattice parameter we find an instability similar to that reported by Liu and co-workers. [35] We repeated the calculation using the local density approximation (LDA), and also find a weak instability. With the experimental lattice parameter we obtain an energy gain upon $R$-point octahedral rotation about [001] of 5 meV per formula unit with the PBE GGA and 4 meV per formula unit with the LDA. These numbers are comparable to the values obtained for BaZrO$_3$ (3 meV per formula unit in the LDA) with the same approach. [49] BaZrO$_3$ is a cubic perovskite material that shows no temperature dependent structural transition but for which a small tilt instability exists in density functional calculations. [49–51] It will be of interest to measure the low temperature structure of BaSnO$_3$ to determine if it is similar to BaZrO$_3$ or if there is a distortion in accord with density functional calculations. In any case, we find that the calculated band gap of BaSnO$_3$ is only weakly dependent on this distortion. On the other hand we find a strong dependence of the gap on volume. This is shown in Fig. 9, and provides an explanation for the differences between the different compounds.

Turning to the other compounds, for rhombohedral ferroelectric KNbO$_3$ we obtain an indirect gap of 2.99 eV, while for ZnSnO$_3$ we obtain 3.07 eV. Thus the ordering of the gaps is CaSnO$_3$> KTaO$_3$> ZnSnO$_3$ $\approx$ KNbO$_3$> BaSnO$_3$. Considering that the O 2$p$ valence bands in perovskites are often aligned, this would suggest that different choices of composition within these systems could yield different results for which side of an electron doped interface has a 2DEG, i.e. a Sn or a Ta/Nb based 2DEG. However, the large difference between the band gaps of BaSnO$_3$ and CaSnO$_3$ implies a strong sensitivity to structural details, which may play an important role at the interface. Actually, the fact that BaSnO$_3$ has the smallest gap would suggest that, all things being equal, the most favorable cases for a 2DEG on the Sn side would be on substrates with the largest lattice parameters, should such substrates become available.

## Supercells

The properties of perovskite thin films are known to be sensitive to strain, which can be controlled in various ways, most commonly choice of substrate, choices of buffer layers and choices of superlattice stackings. Here we are concerned mainly with the question of whether it is possible to have a 2DEG on either side of the interface.

We started with 4 layers of CaSnO$_3$ with 6 layers of KTaO$_3$. We fixed the in plane perovskite lattice parameter to be 3.99 Å (with a $\sqrt{2} \times \sqrt{2}$ cell), i.e. equal to that of KTaO$_3$, and relaxed the $c$-axis lattice parameter, obtaining 39.08 Å, for the 100 atom supercell. This is a condition that would be appropriate for a film of CaSnO$_3$ grown on a TaO$_2$ terminated KTaO$_3$ substrate. We obtained a 2DEG on the Ta side of the interface in this case, as discussed below. Next we replaced KTaO$_3$ by KNbO$_3$. We kept the same cell parameters as for the KTaO$_3$//CaSnO$_3$ supercell, but relaxed all internal coordinates. This cell also produced a 2DEG at the electron rich interface on the KNbO$_3$ side.

We next replaced Ca by Zn. In this case we investigated a supercell consisting of four layers of KNbO$_3$ and six layers of ZnSnO$_3$. Also, differently from the CaSnO$_3$ case, we relaxed all lattice parameters rather than holding the in-plane lattice parameter fixed. This is a condition appropriate to a thick superlattice. The relaxed lattice parameters were $a = 5.38$ Å, $b = 5.36$ Å, and $c = 41.92$ Å. The effective average in-plane perovskite lattice parameter was 3.80 Å, i.e. 4.8% smaller than the CaSnO$_3$ based supercells. The resulting 2DEG electron gas is found to be in the ZnSnO$_3$ layer, but not at the electron rich interface, but rather at the opposite interface. This is a consequence of the fact that ZnSnO$_3$ is ferroelectric (see below).

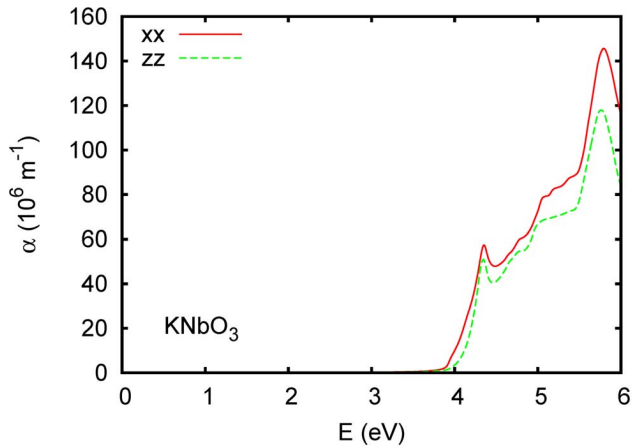

**Figure 8. Calculated absorption spectrum of rhombohedral ferroelectric KNbO$_3$ using the TB-mBJ potential.** Here zz is for polarization along the rhombohedral axis, while xx is perpendicular. Note the different vertical scale from Fig. 2. A Lorentzian broadening of 0.025 eV was applied.
doi:10.1371/journal.pone.0091423.g008

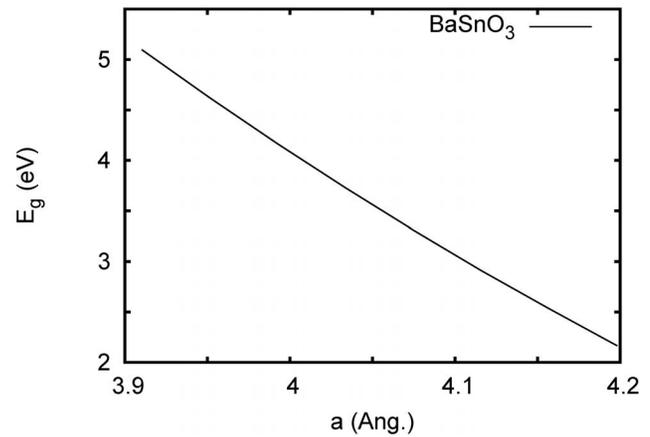

**Figure 9. Calculated band gap of BaSnO$_3$ as a function of lattice parameter with the TB-mBJ potential.**
doi:10.1371/journal.pone.0091423.g009

We also did calculations for four layers of KTaO$_3$ and six layers of ZnSnO$_3$, using the same lattice parameters as the KNbO$_3$//ZnSnO$_3$ supercell, but again relaxing all internal atomic coordinates. Again in this case, we find the 2DEG in the ZnSnO$_3$ at the layer opposite to the electron rich interface. The resulting structures are depicted in Fig. 10. As may be seen, the structures of the CaSnO$_3$ and ZnSnO$_3$ parts of the supercells are more strongly distorted than the KNbO$_3$ and KTaO$_3$. The distortions consist of octahedral tilts and A-site off-centering, the latter being very pronounced for the ZnSnO$_3$ containing supercells.

The conduction band structures for the four supercells described above are shown in Fig. 11 (KNbO$_3$//CaSnO$_3$), Fig. 12 (KTaO$_3$//CaSnO$_3$), Fig. 13 (KNbO$_3$//ZnSnO$_3$), and Fig. 14 (KTaO$_3$//ZnSnO$_3$). In these plots, $\Gamma$-M is the perovskite [100] direction, while $\Gamma$-X is the perovskite [110], which is folded at the mid-point due to the $\sqrt{2} \times \sqrt{2}$ supercell construction. The Fermi energies are fixed by the electron count.

Pairs of bands, in which one member disperses upwards from $\Gamma$ and the other band is weakly dispersive in the $\Gamma$-M direction, come from Nb/Ta layers, specifically $d_{xz}$ and $d_{yz}$ bands in the $t_{2g}$ manifold. The $d_{xy}$ bands, show dispersion along both in plane directions and are split away from the $d_{xz}$ and $d_{yz}$ bands. The Sn s derived bands are also expected to show similar dispersions in both directions. As shown in the band structures, both the CaSnO$_3$ based supercells show similar behavior. In particular, the lowest energy conduction bands come from the KTaO$_3$ or KNbO$_3$ side of the interface and show the structure mentioned above. A series of bands is seen corresponding to the different Nb or Ta layers, but the lowest band is from the layer immediately adjacent to the electron rich interface as might be anticipated. In both supercells the lowest band at $\Gamma$ is from the $d_{xy}$ orbital. The next lowest is from the $d_{xy}$ band of the second layer and the $d_{xz}/d_{yz}$ bands of the interface and subsequent layers lie above. The calculated effective masses are 0.45 $m_e$ and 0.48 $m_e$ for the KTaO$_3$ and KNbO$_3$ cases, respectively.

The ordering at the $\Gamma$ point CBM is a natural ordering for the $t_{2g}$ manifold at an interface. This is because hopping is disrupted in the direction perpendicular to the interface, but not in plane. Therefore the width of the $d_{xy}$ bands is nominally unaffected by the interface, while the $d_{xz}$ and $d_{yz}$ have their width reduced. If the center of the bands is at the same position then the CBM will be formed from the wider band, i.e. the $d_{xy}$ band. The other ingredient that can affect the band ordering is the crystal field.

Crystal field splittings in transition metal oxides come mainly from hybridization between metal d orbitals with the ligand p states, in this case, $\pi$ interactions involving the Nb or Ta $t_{2g}$ orbitals with the O p orbitals, so that the conduction bands are nominally $t_{2g}$ - $p\pi$ antibonding states. Therefore the $d_{xy}$ band could be shifted up in energy if the in-plane O atoms were moved closer (or alternatively if the c-axis O were moved away). However, because the $t_{2g}$ manifold is governed by $\pi$ interactions as opposed to the $\sigma$ interactions in the $e_g$ manifold these effects are relatively weaker.

Turning to the ZnSnO$_3$ containing supercells, the situation is opposite. We find that the CBM takes Sn s character, i.e. it is on the ZnSnO$_3$ side of the interface. As in the case of the CaSnO$_3$ supercells, one sees a series of similar bands going up in energy from the CBM, but these are s derived bands from different ZnSnO$_3$ layers, rather than $t_{2g}$ bands. This is readily seen from the near isotropic dispersion away from $\Gamma$ and is confirmed by projections of the bands (not shown). Importantly, the lowest band, which forms the CBM is not from the Sn s states adjacent the electron rich interface. Instead it is from the opposite, neutral interface. This indicates a ferroelectric tendency in the ZnSnO$_3$ similar to that in the bulk. In other words a state with an effective polarization of zero, which would place the 2DEG at the electron rich interface, is less stable than a state with the polarization fixed at the interface planar charge density (50 $\mu$C/cm$^2$), which then shifts the 2DEG to the opposite interface (note that this is a constrained situation, so it does not make sense to discuss a bulk polarization).

This polarization in the two ZnSnO$_3$ containing supercells is seen in the structures (Fig. 10). Specifically the cations on both sides of the interface are visibly displaced from the centers of cages formed by the coordinating O atoms. The direction of this displacement, which is the direction of the ferroelectric polarization is shown by the arrows. As seen, the polarization in the ZnSnO$_3$ points away from the charge balanced mixed interface and towards the charge imbalanced interface where the electrons comprising the 2DEG originate, consistent with the above discussion. The polarization in the KNbO$_3$ and KTaO$_3$ parts of the supercell, which do not contain charge carriers in these cells points in the same direction as that of the ZnSnO$_3$.

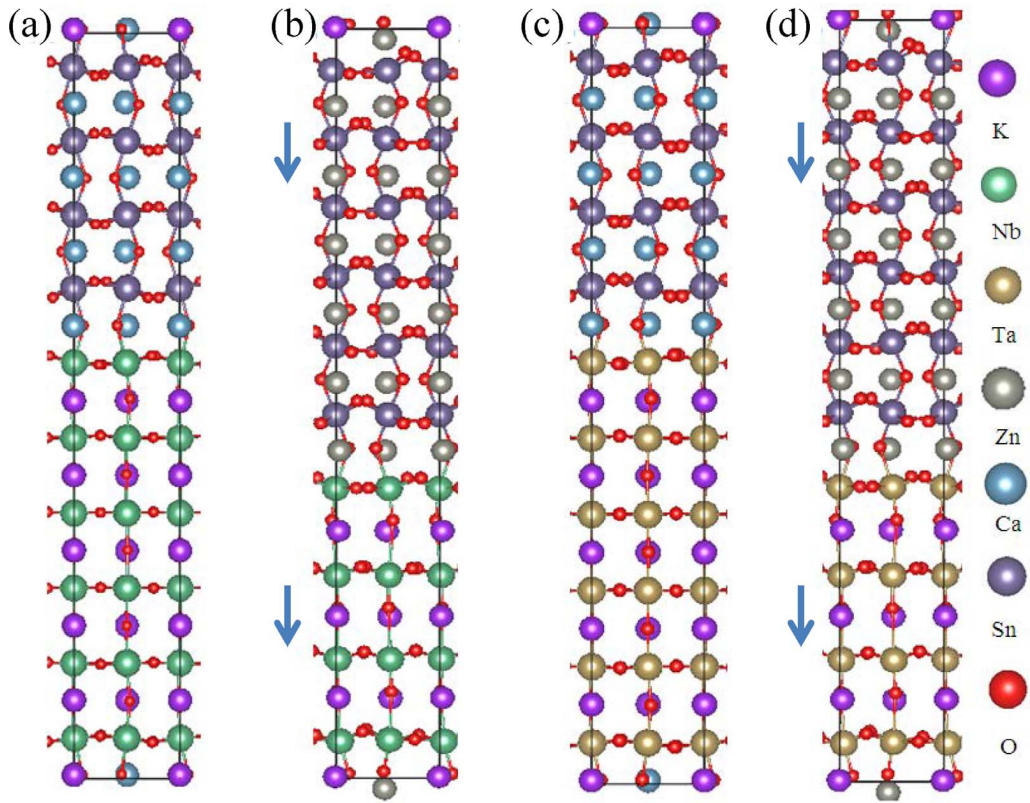

**Figure 10. Structures of supercells, left to right (a) KNbO$_3$//CaSnO$_3$, (b) KNbO$_3$//ZnSnO$_3$, (c) KTaO$_3$//CaSnO$_3$, and (d) KTaO$_3$//ZnSnO$_3$.** As seen, the structures containing ZnSnO$_3$ have noticeable cation offcentering in the coordinating O cages in both the ZnSnO$_3$ and K(Nb,Ta)O$_3$ parts of the supercells. This corresponds to a ferroelectric polarization. The direction of this is indicated by the arrows on the left of the corresponding structure figures.
doi:10.1371/journal.pone.0091423.g010

We note that a related discussion has been presented by Wang, Niranjan and co-workers, who considered 2DEGs at ferroelectric interfaces. [52,53] They predicted that these may be electrically switchable, with important implications for device applications. Based on our results this could also be the case here.

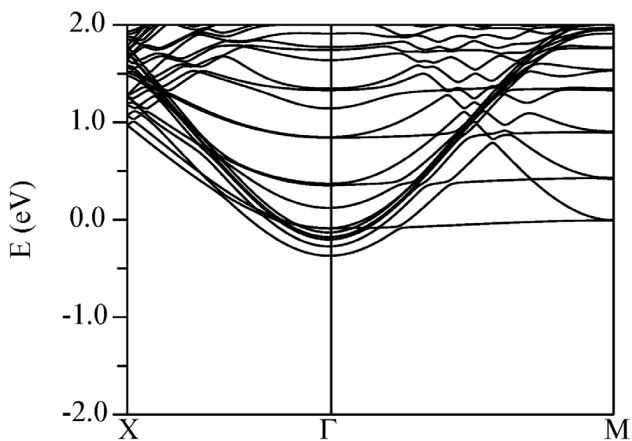

**Figure 11. Calculated conduction band structure of the relaxed (KNbO$_3$)//(CaSnO$_3$) supercell (see text) using the TB-mBJ potential.** The Fermi energy is at 0 eV.
doi:10.1371/journal.pone.0091423.g011

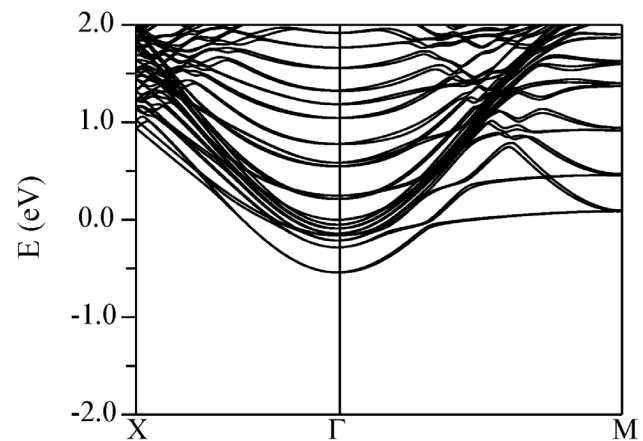

**Figure 12. Calculated conduction band structure of the relaxed (KTaO$_3$)//(CaSnO$_3$) supercell (see text) using the TB-mBJ potential.** The Fermi energy is at 0 eV. Spin orbit is included.
doi:10.1371/journal.pone.0091423.g012

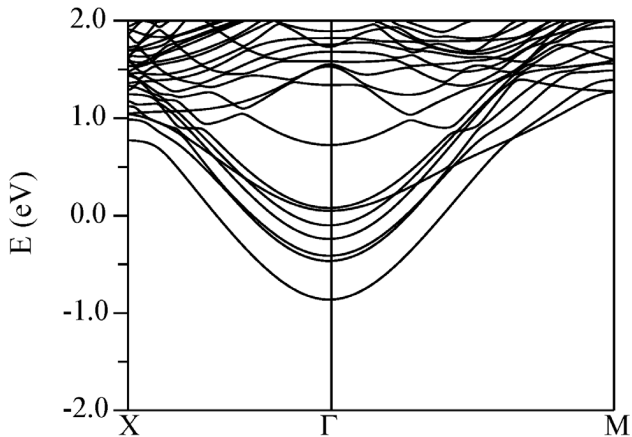

**Figure 13. Calculated conduction band structure of the relaxed (KNbO$_3$)//(ZnSnO$_3$) supercell (see text) using the TB-mBJ potential.** The Fermi energy is at 0 eV.
doi:10.1371/journal.pone.0091423.g013

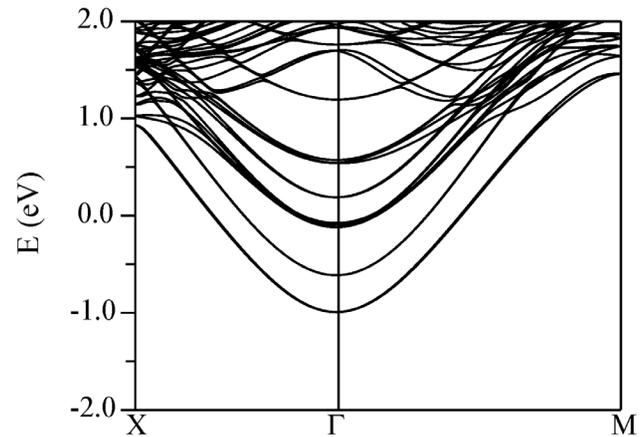

**Figure 14. Calculated conduction band structure of the relaxed (KTaO$_3$)//(ZnSnO$_3$) supercell (see text) using the TB-mBJ potential.** The Fermi energy is at 0 eV. Spin orbit is included.
doi:10.1371/journal.pone.0091423.g014

This may be a particularly interesting form of 2DEG. In particular, because of the ferroelectricity it is shifted away from the charged interface, which may then be viewed as a form of $\delta$-doped 2DEG, i.e. a 2DEG that is in a different physical location than the source of the doping that produces it. It may be then that this could be a particularly high mobility 2DEG if clean, smooth, low defect density capping layers can be grown in practice. The effective mass of the CBM was the same to two figures for both ZnSnO$_3$ supercells and was 0.37 $m_e$, i.e. $\sim$20% lighter than the KTaO$_3$/KNbO$_3$ 2DEGs in the CaSnO$_3$ based supercells.

The calculations above are for highly idealized systems. In practice, adding 1/2 $e$ per unit cell is extreme. As seen in the band structure plots, the Fermi energy is more than 1 eV higher than the CBM for the ZnSnO$_3$ cases. While it is known that these Sn compounds can be heavily doped $n$-type, in general the chemical stability of a phase decreases as one dopes away from the ideal valence. Specifically, one may expect that the energy of defects that compensate the excess charge will decrease as the Fermi energy is raised and so it may very well be that the realizable carrier density is significantly lower than the idealized cases considered here. Electronic reconstructions have also been discussed in the case of SrTiO$_3$/LaAlO$_3$, [54] although it should be noted that this is hard to reconcile with the high mobilities observed in this system. Furthermore, it may be expected that the 2DEG will be sensitive to the details of the structure. This is particularly so for the case where one has ZnSnO$_3$, as in general ferroelectricity is very sensitive to strain.

However, we expect the qualitative features to remain: (1) depending on details it is possible to have 2DEGs in either the transition metal side or the stannate side of the interfaces; (2) the stannate based 2DEGs have lower effective mass than the transition metal based 2DEGs; and (3) ferroelectricity can shift the 2DEG away from the electron rich interface (presumably this can happen for the ZnSnO$_3$, but may also (depending on details such as the strain) occur for strained CaSnO$_3$ films or other similar films such as SrSnO$_3$ or BaSnO$_3$); in this case observation of the 2DEG will depend critically on the termination of the film, e.g. the capping layer. Also, it should be noted that the ordering of the band gaps of the bulk compounds by itself does not predict predict on which side of the interface the 2DEG is found, showing that the details of the interfaces and strain states are important and perhaps can be used to tune these systems.

## Discussion

First principles calculations have been used to study the range of possible behavior for electron rich interfaces between K(Nb,Ta)O$_3$ and (Ca,Zn)SnO$_3$. We find that depending on details it is possible to produce 2DEGs on either side of the interface. The 2DEGs on the stannate side have lower effective mass. A complicating factor is the interplay with ferroelectricity, which can shift the 2DEG away from the interface when it occurs on the stannate side. This suggests that there may be a very interesting, and perhaps electrically controllable interplay between strain, ferroelectricity and the 2DEG electronic properties in these interfacial systems. This also implies that the possibility of ferroelectricity shifting the 2DEG away from the interface should be kept in mind when interpreting experimental results on interfaces with these stannates, as depending on the conditions it may result in non-observation of an expected 2DEG. This is a general issue for 2DEGs produced by interfaces involving ferroelectric materials. It will be of considerable interest to investigate the behavior of these and related electron rich interfaces between stannates (e.g. ZnSnO$_3$, CaSnO$_3$, SrSnO$_3$ or BaSnO$_3$) and transition metal based compounds (e.g. KNbO$_3$, KTaO$_3$ or even LaTiO$_3$) from an experimental point of view.

## Materials and Methods

We investigate the properties of the interfaces using density functional calculations. The results shown are for 100 atom $\sqrt{2}\times\sqrt{2}\times10$ supercells consisting of layers of K(Ta,Nb)O$_3$ and (Ca,Zn)SnO$_3$ stacked along the [001] direction, with a $\sqrt{2}\times\sqrt{2}$ in-plane structure to allow the possibility of rotation/tilts of the octahedra. Similarly, we used even numbers of layers to avoid artificially blocking octahedral tilts that may be important for properly describing the structure. In most calculations we used six layers of K(Ta,Nb)O$_3$ and four layers of CaSnO$_3$, or four layers of K(Ta,Nb)O$_3$ and six layers of ZnSnO$_3$. Each case constructed had two interfaces at $A$O layers. For one interface we used a mixture of the two cations to construct a charge neutral interface, i.e. KCaO$_2$ or KZnO$_2$ (balancing the compounds on either side of the layer), while at the other we constructed an electron rich charge imbalanced interface, i.e. Ca$_2$O$_2$ or Zn$_2$O$_2$. The excess charge of 1$e$ per two perovskite unit cells amounts to 50 $\mu$C/cm$^2$. This is similar to the polarization of bulk ZnSnO$_3$, which has a reported value of 59 $\mu$C/cm$^2$. [55] This choice of supercell results in non-centrosymmetric cells, which are not usually selected because the absence of inversion symmetry slows density functional calculations. On the other hand, it allows the layers to be truly polar, which turns out to be important in the ZnSnO$_3$ containing supercells, as discussed below. As mentioned, we also used a $\sqrt{2}x\sqrt{2}$ in plane structure to allow octahedral rotations and tilts, i.e. the choices made allow the main classes of perovskite instabilities, arbitrary polar off-centering as well as octahedral tilts.

The atomic positions were determined by structure relaxation using the standard generalized gradient approximation (GGA) of Perdew, Burke and Ernzerhof. [56] No symmetry was imposed in these relaxations. We then calculate the electronic structures using the modified Becke-Johnson potential functional developed by Tran and Blaha, [57] which we denote TB-mBJ. This potential, unlike standard GGA functionals, which are designed to reproduce total energies, gives band gaps in reasonable accord with experiment for many simple semiconductors and insulators [58–61].

Initial relaxations were done using the VASP code, [62,63] while final relaxations of the atomic positions and electronic structures were done using the linearized augmented planewave (LAPW) method [64] as implemented in the WIEN2k code. [65] Well converged basis sets were used. In particular, we used local orbitals with the standard LAPW augmentation to accurately include semicore states (i.e. the LAPW+LO method). [66] The standard LAPW augmentation was used for all other states as well, except that the augmented planewave plus local orbital (APW+lo) basis [67] was used for the O 2$p$ states to accelerate convergence. In all cases we used metal sphere radii substantially larger than for the O. This was done to give good convergence for the metal atoms using the standard LAPW+LO method, which gives an accurate treatment in cases with semicore states as is required for early transition elements such as Nb and Ta as well as for the $d$ states of Zn (see the discussion in Refs. [66] and [68]). For O better convergence was obtained using the APW+lo method for the 2$p$ states, since there is not semicore in this case and the small sphere radius makes linearization errors negligibly small. The sphere radii used in the supercells were 2.1 Bohr for K and Ca, 2.25 Bohr for Sn, 1.85 Bohr for Ta, 2.0 Bohr for Nb, 2.05 Bohr for Zn and 1.4 Bohr for O except for the KNbO$_3$/CaSnO$_3$ supercell for which an O radius of 1.45 Bohr was used.

We used dense Brillouin zone samplings, which were needed especially because of the dispersive nature of the Sn s-derived conduction bands. These amounted to at least 8×8 in-plane meshes for the $\sqrt{2} \times \sqrt{2} \times 10$ supercells. Optical properties, shown for some of the bulk compounds, were calculated using the electric dipole matrix elements via the optical package of the WIEN2k code. For this we used denser three dimensional k-point grids of 16×16×16 for the cubic compounds and similarly dense grids in the folded zones of the non-cubic cases. For the the bulk transport calculation we used a 32×32×32 mesh.

## Acknowledgments

We are grateful for helpful discussions with Bharat Jalan.

## Author Contributions

Conceived and designed the experiments: XFF WTZ XC DJS. Performed the experiments: XFF XC DJS. Analyzed the data: XFF XC DJS. Wrote the paper: DJS.